\documentclass[10pt]{iopart}
\usepackage[utf8]{inputenc}
\expandafter\let\csname equation*\endcsname\relax 
\expandafter\let\csname endequation*\endcsname\relax
\usepackage{amsmath}
\usepackage{amssymb}
\usepackage{siunitx}
\usepackage[pdftex]{graphicx}  
\usepackage{todonotes}
\usepackage[T1]{fontenc}
\usepackage{tabularx}
\usepackage{braket}
\usepackage{comment}
\usepackage{pifont}

\begin{document}

\title[Vibrational state inversion of a BEC]{Vibrational state inversion of a Bose-Einstein condensate: optimal control and state tomography}

\author{Robert B\"ucker,  Tarik Berrada, Sandrine van~Frank, Jean-Fran\c{c}ois Schaff, Thorsten Schumm, and J\"org Schmiedmayer}
\address{Vienna Center for Quantum Science and Technology, Atominstitut, TU Wien, Stadionallee 2, 1020 Vienna, Austria}

\author{Georg J\"ager, Julian Grond, and Ulrich Hohenester}
\address{Institut f\"ur Physik,
  Karl--Franzens--Universit\"at Graz, Universit\"atsplatz 5,
  8010 Graz, Austria}

\ead{schmiedmayer@atomchip.org}

\begin{abstract}
We present theoretical and experimental results on high-fidelity transfer of a trapped Bose-Einstein condensate into its first vibrationally excited eigenstate.
The excitation is driven by mechanical motion of the trap, along a trajectory obtained from optimal control theory.
Excellent agreement between theory and experiment is found over a large range of parameters.
We develop an approximate model to map the dynamics of the many-body condensate wave function to a driven two-level system.
\end{abstract}

\maketitle

\section{Motivation}

The precise control over quantum systems represents a major challenge in modern physics. 
Successful implementation of quantum technologies may lead to the construction of devices such as quantum simulators, quantum cryptography devices, and quantum computers. 
For such applications, one needs to produce arbitrary quantum states, e.g.\ strongly entangled many-body states, or states which are far from thermal equilibrium or the ground state of the system.

In this article we report on highly efficient preparation of a non-classically excited motional state of a Bose-Einstein condensate (BEC), by a modulation of the trapping potential, as obtained from optimal control theory (OCT). 
Fast changes of the potential are routinely used in BEC laboratories, for instance as ways to probe the gas by exciting collective excitations~\cite{Jin1996, Mewes1996}, or to displace the samples for further manipulation. 
Recently, controlled modulations of the trapping potential were achieved in order to quickly displace BECs while keeping them in their ground state. 
These ``shortcuts to adiabaticity''~\cite{Couvert2008, Muga2009, Schaff2011, Schaff2011d,Bason2011} take advantage of the self-similar dynamics of interacting BECs trapped in time-dependent harmonic potentials~\cite{Kagan1996, Castin1996}. 

For more general desired states, like excited stationary states, for which no exact solutions are found, one needs to use numerical methods such as OCT. 
Such approaches were investigated theoretically for the splitting of BECs~\cite{Hohenester2007}, to optimize the transport of atoms in optical lattices for quantum gate operations~\cite{DeChiara2008}, or to amplify number squeezing~\cite{Grond2009}.

\begin{figure}
\centering
\includegraphics{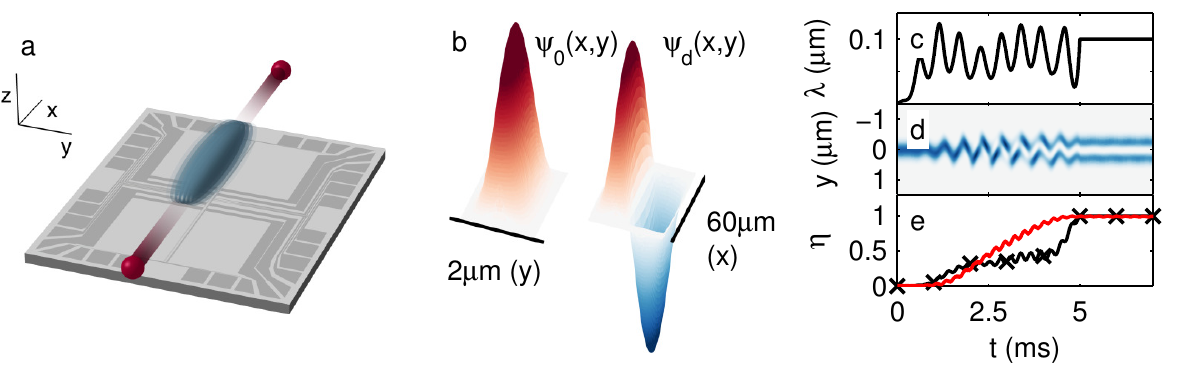}
\caption{Schematic representation of our excitation scheme.
(a) Illustration of the sequence: an elongated Bose-Einstein condensate (blue) is trapped on an atom chip (chip and condensate aspect ratio not drawn to scale).
Mechanical motion along the y-axis pumps the gas into a vibrationally excited state, which decays by directed emission of twin-atom beams along x (red).
(b) Initial and excited condensate wave functions in the (x,y)-plane.
The state $\psi_\mathrm{d}(x,y)$ (desired state) is the first excited state of the Gross-Pitaevskii equation~\eqref{eq:gp} along y.
(c) Trajectory $\lambda(t)$ of the trap minimum along $y$.
(d) Time-dependent density $n(y,t)=|\psi(y,t)|^2$ of the condensate wave function under influence of the excitation process.
(e) Population (simulated) of excited state $\psi_\mathrm{d}(y)$ derived from wave-function overlap (black, with markers) and using the two-mode model as introduced in Sec.~\ref{sec:shake_two_level} (red, solid).}
\label{fig:overview}
\end{figure}

Here, we aim for a vibrational state inversion, where the entire population of the condensate is transferred to the first excited state of its motional degree of freedom.
Such an inverted state can be used as a source for the amplified emission of matter-wave twin beams~\cite{Buecker2011,Buecker2012}, similar to a pumped gain medium in a laser or an optical parametric amplifier~[Fig.~\ref{fig:overview}(a)].f
We start from a condensate in the ground state along the strongly confined (transverse) directions of an elongated trapping potential.
We then use OCT on a controlled displacement of the trap center [transverse ``shaking'' of the cloud, Fig.~\ref{fig:overview}(c)], in order to transfer the BEC to the first antisymmetric stationary state as given by the Gross-Pitaevskii equation (GPE), which is governing the system's dynamics [Fig.~\ref{fig:overview}(b)].
The efficiency of this process is close to 100\%, which corresponds to the desired vibrational inversion of the atomic cloud.
Since a time-dependent harmonic potential (where all energy levels are equidistant) would not allow to transfer to an excited stationary state \cite{Jacak1998, Walls2007,Garcia-Ripoll2001,Bialynicki-Birula2002}, we here use an anharmonic potential~\cite{Khani2009,Jirari2009} generated by a radio-frequency dressed magnetic trap~\cite{Lesanovsky2006}. 
For a non-interacting gas, the final state would simply correspond to all the atoms residing in the first excited eigenstate of the trap (quantum numbers $n_\text{x} = 0, n_\text{y} = 1, n_\text{z} = 0$). 
However, in our many-body wave function inter-atomic interactions are an essential ingredient of the system's dynamics and cannot be neglected in the optimization.
In the experiment they manifest themselves in energy shifts due to the atomic mean-field, and a decay of the excited state by means of inelastic two-body scattering~\cite{Buecker2011,Buecker2012}.

Another aspect we will address is the interpretation of our results beyond a simple comparison of calculated and measured wave function dynamics.
While such a comparison benchmarks the accuracy to which experiments and theory are matched, it provides only limited insight into the nature of the excitation mechanism, and the structure of the quantum state during and after the excitation process.
To this end, we will deduce an approximate description, that allows to map the many-body wave function in a weakly anharmonic confinement to a driven two-level system, where the excitation process corresponds to a $\pi$-pulse that transfers all population to the excited state.

The paper is structured as follows: section~\ref{sec:theory} presents the theoretical description of the problem and the OCT algorithm used to obtain the excitation trajectory of the trap center, section~\ref{sec:xp} details the experimental implementation, and finally, section~\ref{sec:results} discusses the results, with an emphasis on how the behavior of key observables can be captured by a two-level model. 
To our knowledge, this excitation sequence represents the first successful use of OCT for the preparation of exotic many-body states of Bose-Einstein condensates.

\section{Optimal control theory}
\label{sec:theory}


Optimal control theory (OCT) is a mathematical tool that allows to determine an optimal control sequence for a given control problem \cite{peirce:88,rabitz:00}.  In the following we review the basic ingredients of optimal control theory taking the example of the shaking process which brings the condensate from the ground state to its first vibrationally excited eigenstate.  Our analysis closely follows the presentation given in Refs.~\cite{hohenester.pra:07,Buecker2012}.

As will be discussed in Sec.~\ref{sec:xp}, the condensate can be described by a one-dimensional Gross-Pitaevskii equation for the transverse coordinate $y$, along which the condensate is displaced, according to
\begin{equation}\label{eq:gp}
  i\hbar\frac{\partial\psi(y,t)}{\partial t}=\left(
  -\frac{\hbar^2}{2m}\frac{\partial^2}{\partial y^2}+
  V_\lambda(y,t)+g\left|\psi(y,t)\right|^2\right)\psi(y,t)\,.
\end{equation}
Here $m$ is the mass of the $^{87}$Rb atoms and $g$ is a nonlinearity parameter accounting for the repulsive atom-atom interactions \cite{hohenester.pra:07,Grond2009}. The anharmonic confinement potential $V_\lambda(y,t)=V_6(y-\lambda(t),0)$ (see Sec.~\ref{sec:trap}) follows a control parameter $\lambda(t)$, in our case the displacement of the potential minimum, and provides the means for exciting the condensate.  The objective of the control problem can now be formulated as follows.  Let $\lambda_0$ be the control parameter at the initial time $t=0$, and $\lambda_1$ the control parameter at the final time $t=T$ of the control process.  Likewise, we denote the initial ground state of the GPE with $\psi_0(y)$ and the \textit{desired} final wave function (in our case the first excited state of the GPE in the anharmonic trap) with $\psi_\mathrm{d}(y)$.  OCT then seeks for the optimal time variation of $\lambda(t)$ that brings the final wave function as close as possible to the desired state $\psi_\mathrm{d}$.

To gauge the success of the excitation process for a given control field $\lambda(t)$, we define a cost function
\begin{equation}\label{eq:cost}
  J(\psi(T),\lambda)=\frac 12\left[1-\left|\langle \psi_\mathrm{d}|\psi(T)\rangle\right|^2\right]+
  \frac\gamma 2\int_0^T\left[\dot\lambda(t)\right]^2\,\mathrm{d}t\,.
\end{equation}
The first term of the cost function becomes minimal when the final wave function precisely matches the desired wave function, apart from a global (irrelevant) phase.  The second term favors smooth control fields and is needed to make the OCT problem well posed \cite{borzi.pra:02}.  $\gamma$ is a parameter that weights the relative importance of the two control objectives of smooth control fields and of wave function matching. As our experimental implementation allows fast and precise control of $\lambda(t)$, the parameter $\gamma$ can be set such that the control penalization is always much smaller than the first term in Eq.~\eqref{eq:cost}. OCT is now seeking for an ``optimal control'' that minimizes the cost function $J(\psi(T),\lambda)$, under the condition that the final wave function $\psi(T)$ has to be obtained from the Gross-Pitaevskii equation of Eq.~\eqref{eq:gp} with the initial wave function $\psi_0(y)$.  To turn this constrained minimization problem into an unconstrained one, within the OCT framework one introduces a Lagrange function
\begin{displaymath}\label{eq:lagrange}
  L(\psi,p,\lambda)=J(\psi,\lambda)+ \Re\mbox{e}\int_0^T\Biggl< p\Biggr|
  i\hbar\frac{\partial\psi}{\partial t}-\left(
  -\frac{\hbar^2}{2m}\frac{\partial^2}{\partial y^2}+
  V_\lambda+g\left|\psi\right|^2\right)\psi\Biggr>\,\mathrm{d}t\,,
\end{displaymath}
where the adjoint function $p(y,t)$ acts as a generalized Lagrange parameter.  Here and in the following we will, for the sake of brevity, often omit parameters $y$ and $t$. At the minimum of $J(\psi,\lambda)$ the Lagrange function has a saddle point, where all three derivatives $\delta L/\delta\psi$, $\delta L/\delta p$ and $\delta L/\delta\lambda$ must vanish.  Performing the usual functional derivatives, we obtain after some variational calculation the following optimality system:
\begin{subequations}\label{eq:optimality}
\begin{eqnarray}
i\hbar\frac{\partial\psi}{\partial t}&=&\left(-\frac{\hbar^2}{2m}\frac{\partial^2}{\partial y^2}+V_\lambda+g|\psi|^2\right)\psi
\label{eq:oct.forward}\\
i\hbar\frac{\partial p}{\partial t}&=&\left(-\frac{\hbar^2}{2m}\frac{\partial^2}{\partial y^2}+V_\lambda+2g|\psi|^2\right)p+
g\,\psi^2\,p^*
\qquad\label{eq:oct.backward}\\
\hbar \gamma\ddot\lambda&=&-\Re e\,\langle\psi|\frac{\partial V_\lambda}{\partial\lambda}|p\rangle
\label{eq:control}\,,
\end{eqnarray}
\end{subequations}
which has to be solved together with the initial condition $\psi(0)=\psi_0$, as well as with the constraints on the control field $\lambda(0)=\lambda_0$ and $\lambda(T)=\lambda_1$.  To obtain the equation for the adjoint function $p$, we have performed an integration by parts for the term involving the time derivative of $\psi$ prior to working out the functional derivative $\delta L/\delta\psi$.  This procedure gives, in addition to Eq.~\eqref{eq:oct.backward}, the terminal condition
\begin{equation}\label{eq:terminal}
  i\hbar\,p(y,T)=-\langle\psi_\mathrm{d}|\psi(T)\rangle\,\psi_\mathrm{d}(y)\,.
\end{equation}
Quite generally, the Lagrange parameter determines the sensitivity of the system with respect to the external control.  In our case, the dynamic equation \eqref{eq:oct.backward} describes the propagation of fluctuations around the Gross-Pitaevskii solution and is closely related to the usual Bogoluibov-de~Gennes  equations \cite{leggett:01}.

In most cases of interest it is impossible to guess $\lambda(t)$ such that Eqs.~(\ref{eq:optimality}a--c) are fulfilled simultaneously, and one has to employ a numerical solution scheme.  Suppose that $\lambda(t)$ is some guess for a viable control field.  We can now solve Eq.~\eqref{eq:oct.forward} forward in time to obtain the final wave function $\psi(T)$, which, in turn, allows us to compute the adjoint function $p(T)$ from Eq.~\eqref{eq:terminal}. In the ensuing step, the time evolution of $p(t)$ is solved backwards in time.  Since $\lambda(t)$ is not the optimal control, Eq.~\eqref{eq:control} is no longer fulfilled.  However, the functional derivative
\begin{equation}\label{eq:grad.L}
  \frac{\delta L}{\delta\lambda}=-\gamma\ddot\lambda-\Re\mbox{e}\langle\psi|
  \frac{\partial V_\lambda}{\partial\lambda}|p\rangle
\end{equation}
provides us with a search direction for $\lambda(t)$.  Adding a fraction of $\delta L/\delta\lambda$ to $\lambda(t)$ leads to a control that performs better and brings the final wave function $\psi(T)$ closer to the desired one.  The improved control field is then used in the next iteration.  In our simulations we typically perform a time discretization of the interval $[0,T]$ and use a generic optimization routine, such as the nonlinear conjugate gradient \cite{press:02} or a quasi-Newton method, together with  Eq.~\eqref{eq:grad.L} for computing the appropriate search directions.  One shortcoming of Eq.~\eqref{eq:grad.L} is that in general $\delta L/\delta\lambda$ does not vanish at the boundary points of the time interval, although the control field is fixed to the values of $\lambda_0$ and $\lambda_1$ there.  To overcome this problem, one rewrites the penalization term of the control field $(\gamma/2)\,\bigl (\dot\lambda, \dot\lambda\bigr)_\mathrm{{L^2}}$ as $(\gamma/2)\,\bigl (\lambda, \lambda\bigr)_\mathrm{{H^1}}$, where the definition of the $\mathrm{H^1}$ inner product is $(u,v)_\mathrm{{H^1}}=(\dot u,\dot v)_\mathrm{{L^2}}$ \cite{Grond2009b}.  It is important to realize that this different norm does neither affect the value of the cost function nor the Gross-Pitaevskii or adjoint equations.  However, it does affect the equation for the control field in case of a non-optimal $\lambda(t)$, which now satisfies a Poisson equation
\begin{equation}\label{eq:searchh1}
  -\frac{\mathrm{d}^2}{\mathrm{d}t^2}\frac{\delta L}{\delta\lambda}=-\gamma\frac{\mathrm{d}^2\lambda}{\mathrm{d}t^2}-\Re\mbox{e}
  \bigl<\psi\bigr|\frac{\partial V_\lambda}{\partial\lambda}\bigl|p\bigr>\,.
\end{equation}
The advantages of Eq.~\eqref{eq:searchh1} are that the boundary conditions for $\lambda(t)$ are automatically fulfilled and that changes due to large values of the second term on the right-hand side are distributed, through the solution of the Poisson equation, over the whole time interval.  In all our OCT calculations we use Eq.~\eqref{eq:searchh1} instead of Eq.~\eqref{eq:grad.L}.


\begin{figure}
\centering
\includegraphics[width=\textwidth]{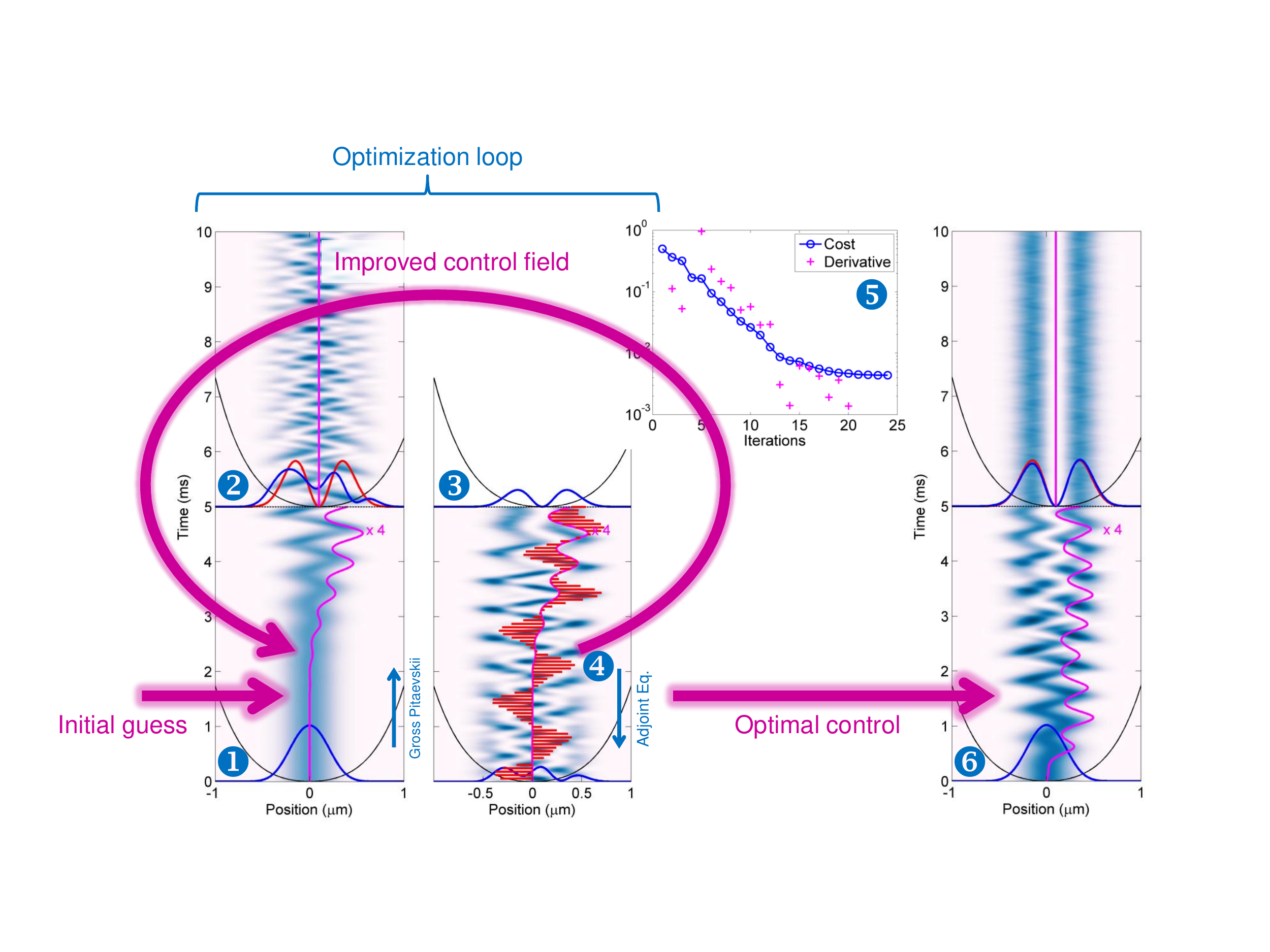}
\caption{Schematics of the OCT optimization loop, which starts with an initial guess for the control field $\lambda(t)$ associated with the displacement of the minimum of the confinement potential.  First, the Gross-Pitaevskii equation with the \ding{172} initial condensate wave function $\psi_0(y)$ is solved forwards in time, to obtain \ding{173} the final wave function $\psi(y,T)$ at the terminal time $T=5$ ms of the control process, which in general deviates significantly from the desired, first excited wave function $\psi_\mathrm{d}(y)$.  The density plots in the different panels report the time evolution of the square moduli of the different functions.  From the knowledge of $\psi(y,T)$ and $\psi_\mathrm{d}(y)$ we can compute the \ding{174} terminal value of $p(y,T)$ via Eq.~\eqref{eq:terminal}, and solve the adjoint equation \eqref{eq:oct.backward} backwards in time \ding{175}, to finally come up with a new search direction for the optimal control field [Eq.~\eqref{eq:searchh1}] that is used in the next iteration of the optimization loop.  The solid lines superimposed on $\lambda(t)$ in the panel of the adjoint equation depict the search directions.  The inset \ding{176} shows how the cost and derivative for a given control decrease with increasing iterations, until \ding{177} an optimal control is obtained. Here $\lambda(t)$ (magnified by a factor 4) steers the system from $\psi_0(y)$ to the desired wave function at the terminal time of the control process.}
\label{fig:oct}
\end{figure}

Our OCT implementation relies on a numerical optimization routine and a differential equation solver.  As for the optimization routine, one can use any generic code that, starting from some initial guess for the control field, requires a function value (the cost function) together with the derivative of the evaluated function $\delta L/\delta\lambda$ to compute a new, improved $\lambda(t)$.  When using the $\mathrm{H^1}$ norm of Eq.~\eqref{eq:searchh1} one must ensure that all inner products in the generic code are evaluated as $(u,v)_\mathrm{{H^1}}$ rather than $(u,v)_\mathrm{{L^2}}$.  In general we observed the best performance for the quasi-Newton BFGS optimization \cite{bertsekas:99}, which outperforms the nonlinear conjugate gradient method for larger number of iterations in the optimization loop.  As for the differential equation solver, we usually employ a split operator technique \cite{hohenester.pra:07} because of its robustness and simplicity. 

The OCT optimization is schematically depicted in Fig.~\ref{fig:oct}.  One starts with some initial guess for the control field.  In general, the outcome of the OCT loop does not depend critically on the initial $\lambda(t)$ and one can use any reasonable guess, such as in our case some interpolating function between the boundary values of $\lambda_0=0$ and $\lambda_1=0.1\,\mu$m at the terminal time $T=5$ ms.  Next, \ding{172} the condensate wave function $\psi(y,0)=\psi_0(y)$ is set to the ground state $\psi_0(y)$ of the anharmonic trap, including the nonlinear term of the Gross-Pitaevskii equation \cite{hohenester.pra:07}, and Eq.~\eqref{eq:oct.forward} is solved forwards in time to obtain \ding{173} the terminal wave function $\psi(y,T)$.  For the initial guess of the control, $\psi(y,T)$ differs significantly from the desired, first excited state $\psi_d(y)$ of the anharmonic trap, which has a node in the middle, as can be also inferred from the ensuing time evolution where the trap displacement is held constant.  From Eq.~\eqref{eq:terminal} we can compute the \ding{174} terminal condition for $p(t)$, and \ding{175} solve the adjoint equation \eqref{eq:oct.backward} backwards in time.  Finally, the knowledge of the complete history of $\psi(y,t)$ and $p(y,t)$ allows us to compute the new search direction through Eq.~\eqref{eq:searchh1}, and to pass this direction to the optimization routine which will come up with a new, improved $\lambda(t)$, which can be used in the next iteration of the optimization loop.

In the inset \ding{176} of Fig.~\ref{fig:oct} we show how the cost function $J(\psi,\lambda)$ and the derivative measure $|\delta L/\delta\lambda|$ evolve with increeasing iterations.  Note that the ``optimal control'' corresponds to a minimum of the control landscape, associated with a derivative equal to zero, but it is generally not guaranteed that also the cost is small there.  However, there are indications that under quite broad conditions the OCT loop will come up with a $\lambda(t)$ that fulfills the control objective of wave function matching almost perfectly \cite{rabitz:04}.  In our simulations we typically stop after a given number of iterations or when the derivative has become sufficiently small.  The resulting $\lambda(t)$ sequence is then called the optimal control.  As can be seen from the solution of the Gross-Pitaevskii equation on the in \ding{177}, with this control we closely match the desired wave function at the terminal time, with a fidelity of $|\langle\psi_\mathrm{d}|\psi(T)\rangle|^2 \approx 1-3 \cdot 10^{-3}$. Up to a global phase, the wave function remains stationary for $t > T$.

\section{Experimental implementation}
\label{sec:xp}

The vibrational state control scheme is realized using an ultra-cold Bose gas trapped on an atom chip.
The experimental procedure is very close to that described in~\cite{Buecker2011,Buecker2012}.
In brief, a laser-cooled cloud of Rubidium-87 atoms in the $\ket{F=1,m_\mathrm{F}=-1}$ Zeeman level is loaded into a strongly elongated atom chip wire trap~\cite{Reichel2011,Trinker2008a}.
Using forced evaporative cooling, the gas is brought to a temperature close to quantum degeneracy.
Then, by means of radio-frequency dressing~\cite{Schumm2005a,Lesanovsky2006}, the external confinement along the two tightly trapped axes $y,z$ is deformed from a harmonic to an anharmonic and anisotropic potential (see next section for details).
In the anharmonic trap, further cooling down to a temperature $T$ well below $\SI{50}{nK}$ is performed.
With $N\sim 800$ atoms, the gas is now in a quasi-condensate~\cite{Kheruntsyan2003,Petrov2000} regime, where phase fluctuations prevent true condensation of the matter wave along the longitudinal (elongated) direction $x$.
However, the energy scales correponding to both temperature ($k_\mathrm{B} T \sim h \times \SI{550}{Hz}$) and atom interactions (chemical potential $\mu \sim h \times \SI{600}{Hz}$) are well below the trap level spacing ($\sim h \times \SI{2}{kHz}$) along the transverse directions $y,z$.
Thus, thermal excitations in the transverse degrees of freedom are frozen out, and almost all atoms occupy a single transverse mode; along $y$ and $z$, the system is hence appropriately described by a single condensate wave function.
In this paper, we are mostly concerned with the dynamics along the transverse direction $y$, and hence neglect the longitudinal mode structure of the quasi-condensate.

Having prepared the system in this way, we apply the control sequence, while monitoring the momentum space distribution of the condensate, as will be described in the following sections.

\subsection{Trap preparation}
\label{sec:trap}

In a quantum harmonic oscillator, all states that can be addressed by simple displacement of the potential are quasi-classical coherent states~\cite{Walls2007}.
This statement also holds for a harmonically trapped interacting many-body system, where a quasi-classical collective oscillation at the trap frequency fully decouples from more complex internal dynamics~\cite{Garcia-Ripoll2001,Bialynicki-Birula2002}.
Hence, transferring the condensate population into an excited, stationary state necessitates an \emph{anharmonic} potential along the displacement direction $y$, where the decoupling of collective and internal dynamics breaks down.
Furthermore, to be robust against excitations in the perpendicular direction $z$, anisotropy in the transverse plane of the potential is required, causing a detuning of trap levels between the directions.

\begin{figure}
\centering
\includegraphics[width=10cm]{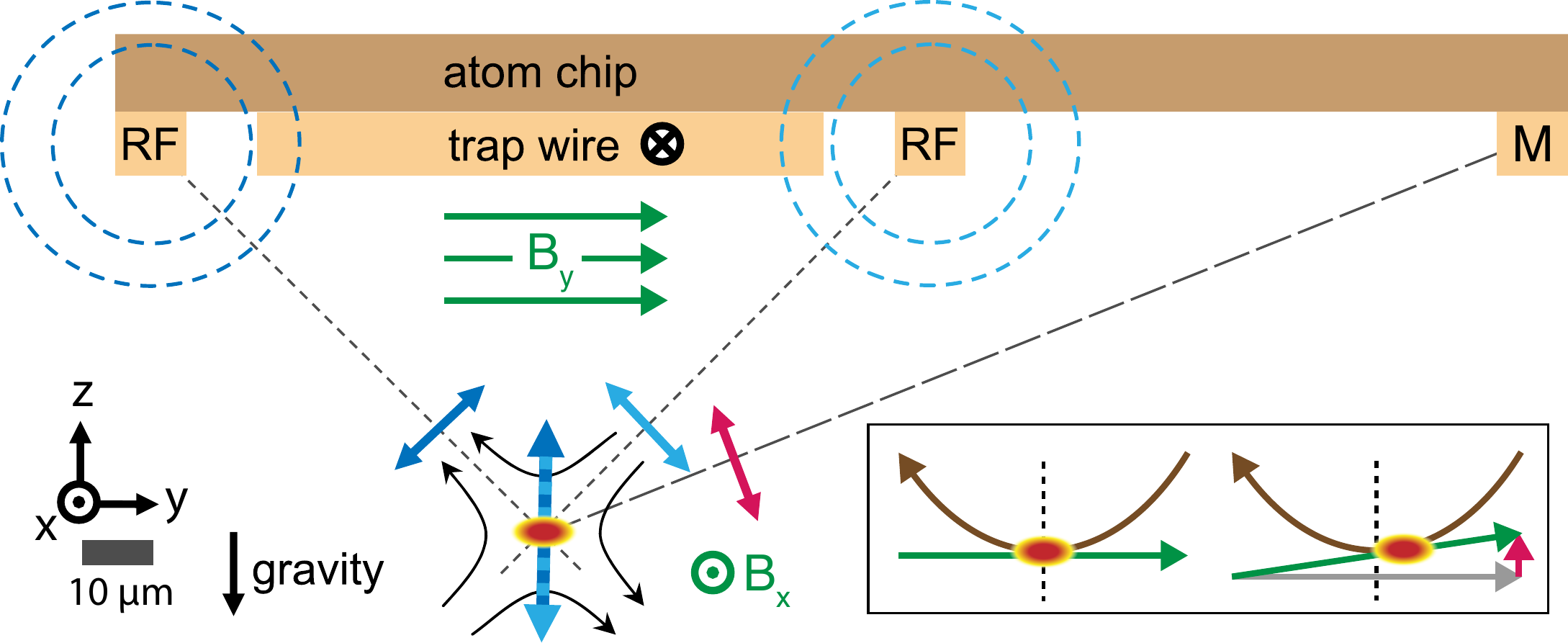}
\caption{(Main figure) Schematic of the atom chip layout (see Ref.~\cite{Trinker2008a} for details).  
The waveguide potential is formed by the current through the trap wire along $-x$ and a static bias field $B_\mathrm{y}$, adding up to quadrupole field (bent arrows).
An external offset field along $B_\mathrm{x}$, perpendicular to the figure plane, defines the Larmor frequency at the trap minimum (Ioffe-Pritchard field configuration).
On a separate chip layer, currents in broad wires along y (not shown) provide weak longitudinal confinement.
The radio frequency dressing currents are applied to wires (RF) in parallel to the trapping wire, leading to a RF field along z (blue arrows).
The resulting anisotropic transverse potential is shown as ellipse in the center of the quadrupole.
Finally, the modulation of the trap position is accomplished by a current in an auxiliary wire (M), leading to a magnetic field, aligned at $\sim 19^\circ$ with respect to the z axis (red arrow).
(Inset) Field configuration for trap position modulation.
The transverse trap position is defined by cancellation of the chip wire field (brown) and the bias field (green).
Adding a weak field along z (red) tilts the bias field slightly, leading to a horizontal shift of the trap minimum. 
}
\label{fig:wires}
\end{figure}

\begin{figure}
\centering
\includegraphics{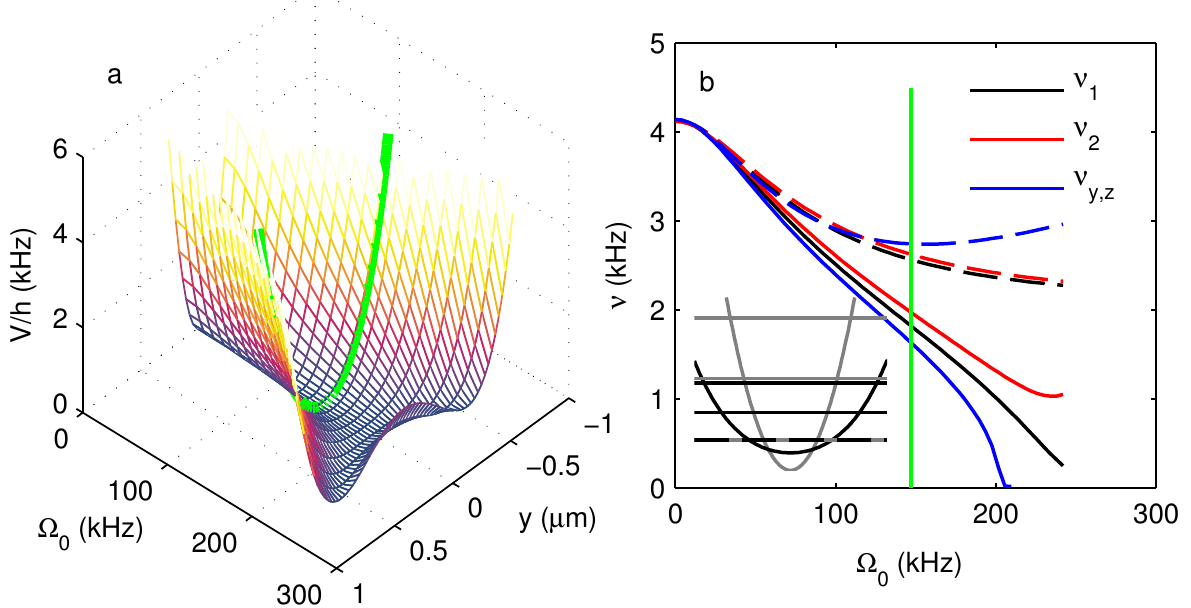}
\caption{
Effects of RF dressing on the transverse trapping potential.
(a) Potential along the y (displacement) direction, as a function of RF Rabi frequency.
The detuning is $\Delta_0 = \SI{-55}{kHz}$.
At dressing strengths above $\Omega_0 \sim \SI{180}{kHz}$, splitting of the single potential into a double well occurs.
(b) Shift of single-particle trap levels vs. dressing strength.
Solid and dashed lines correspond to perpendicular ($y$) and parallel ($z$) directions with respect to the RF polarization, respectively.
Blue: frequency of harmonic part, as defined in Eq.~\eqref{eq:potential6_initial}.
Black, red: first and second level spacing of single-particle eigenstates.
Inset: initial (grey) and dressed (black) potential, each with their first three energy levels.
The green lines in both panels mark the setting used for the experiments.}
\label{fig:potential}
\end{figure}

Initially, the Ioffe-Pritchard field configuration as created by the chip wires (plus external offset fields, see Fig.~\ref{fig:wires}) is rotationally symmetric, and provides harmonic trapping along the transverse directions $y,z$.
For the parameters chosen in our experiment, the transverse trap frequency is $\nu_0=\SI{4.1}{kHz}$ in both directions, whereas the longitudinal frequency is of the order of $\SI{30}{Hz}$.
To introduce anharmonicity and anisotropy, we apply radio-frequency dressing~\cite{Zobay2004,Lesanovsky2006,Schumm2005a}.
Using two chip wires running in parallel to the trapping wire as antennae, the atoms are irradiated by a radio-frequency (RF) near field with linear magnetic polarization, which is red-detuned by tens of $\si{kHz}$ with respect to the atomic Larmor frequency due to the static magnetic field at the trap center.
The RF field adiabatically mixes the Zeeman levels of the $F=1$ hyperfine manifold, coupling them to dressed states. 
This gives rise to an energy shift that depends on detuning from the Larmor frequency $\Delta(\mathbf{r})$ and coupling strength (Rabi frequency) $\Omega(\mathbf{r})$.
Both quantities are position-dependent, the latter because of the changing RF polarization with respect to the local magnetic field that modulates the coupling strength.
In rotating-wave approximation~\cite{Lesanovsky2006}, the resulting potential landscape up to a constant is given by:
\begin{align*}
V(\mathbf{r})/h = \sqrt{\Omega(\mathbf{r})^2 + \Delta(\mathbf{r})^2}.
\end{align*}
The dressing is most effective along the direction perpendicular to the RF polarization; in our case, applying a polarization along the vertical axis $z$ leads to a deformation mostly along $y$.
In Fig.~\ref{fig:potential}(a), the potential along $y$ is shown as a function of dressing strength, expressed as coupling $\Omega_0$ near the trap center.
At sufficiently strong coupling, splitting of the potential into a double well occurs, which is the typical application of RF-dressed potentials~\cite{Schumm2005a,Hofferberth2007b,Hofferberth2008,Betz2011,Gring2012}.
However, at lower coupling, this technique also allows for the introduction of anharmonicity and anisotropy to a single trap, as needed for our scheme.
In the experiment, we apply a RF field of $\sim\SI{0.84}{G}$ peak-to-peak amplitude, leading to a coupling $\Omega_0 = \SI{147}{kHz}$, at a frequency red-detuned by $\Delta_0 = -\SI{54}{kHz}$ with respect to the Larmor frequency near the trap minimum (\SI{824}{kHz}).

The resulting potential is shown as a green line in Fig.~\ref{fig:potential}(a).
Even though the rotating-wave approximation holds well for the used dressing strength~\cite{Hofferberth2007a}, the high sensitivity of the excitation protocol to the exact potential shape calls for an exact calculation by means of a Floquet analysis \cite{Shirley1965}.
Along two transverse directions the result can be approximated by a sixth-order polynomial of the form 

\begin{align}
\label{eq:potential6_initial}
V_6(y,z)/h &= \frac{\nu_\mathrm{y}}{2} \left(\frac{y}{l_\mathrm{y}}\right)^2 + \sigma_\mathrm{y} \left( \frac{ y }{ l_\mathrm{y} } \right)^4 + \xi_\mathrm{y} \left(\frac{y}{l_\mathrm{y}}\right)^6\\
& + \frac{\nu_\mathrm{z}}{2} \left(\frac{z}{l_\mathrm{z}}\right)^2 + \sigma_\mathrm{z} \left( \frac{ z }{ l_\mathrm{z} } \right)^4 + \xi_\mathrm{z} \left(\frac{z}{l_\mathrm{z}}\right)^6.
\end{align}
In this expression, the lengths $l_\mathrm{y,z} = \sqrt{h/(m \nu_\mathrm{y,z})}/(2\pi)$ correspond to the characteristic length of the harmonic part.
The parameters are given by:
\begin{align}
\nu_\mathrm{y} &= \SI{1655}{Hz}; & \nu_\mathrm{z} &= \SI{2751}{Hz}\\
\sigma_\mathrm{y} &= \SI{78.2}{Hz} \nonumber & \sigma_\mathrm{z} &= \SI{-69.6}{Hz} \\
\xi_\mathrm{y} &= \SI{-0.96}{Hz} \nonumber & \xi_\mathrm{z} &= \SI{9.1}{Hz} \\
l_\mathrm{y} &= \SI{265}{nm} & l_\mathrm{z} &= \SI{206}{nm}. \nonumber
\end{align}
Along $y$, the sixth-order term $\xi_\mathrm{y}$ is negligibly small, and the description reduces to a Duffing oscillator~\cite{Amore2005}.

By solving the Schrödinger equation, the single-particle trap levels of the dressed potential can be obtained. 
The first two level spacings $\nu_{1,2}$ along $y$ and $z$ are shown in Fig.~\ref{fig:potential}(b).
For the used parameters (as marked by a green line), the initial degeneracy of the level spacings is lifted, and we obtain the excitation energies (zero-point energy subtracted) $\left[E_{10}, E_{20}, E_{01},E_{02},E_{11}\right]/h = \left[1.84, 3.83, 2.58, 5.21\right]~\si{kHz}$ with $E_{ij}$ denoting the $i$-th and $j$-th state along $y$ and $z$, respectively.
The relevant level spacings along $y$ are given by $\nu_1= \SI{1.84}{kHz}, \nu_2 = \SI{1.99}{kHz}$, the first level spacing along $z$ is $\nu_\mathrm{z}=\SI{2.58}{kHz}$.
From the corresponding eigenfunction along the $z$ direction, and a Thomas-Fermi approximation~\cite{Pethick2002} of the longitudinal profile for $N=800$ atoms, we can estimate the coupling constant\footnote{Note, that we normalize the wave function to 1, not $N$, in Eq.~\eqref{eq:gp}. Hence, $g$ incorporates the atom number.} in Eq.~\eqref{eq:gp} by averaging as $g = h \times\SI{300}{Hz \micro m}$~\cite{Salasnich2002}.

In the experiment, characterization of the initial harmonic trap is straightforward, using radio-frequency spectroscopy and observation of collective oscillations.
On the other hand, confirming the (calculated) parameters of the anharmonic dressed trap with sufficient accuracy is difficult.
Instead, we optimize the experimental control parameters (RF field strength and detuning) directly, by comparing the response to the control ramp to that determined numerically using those trap parameters.
Along the longitudinal $x$-axis, the harmonic trap frequency $\nu_\mathrm{x}=\SI{16.3}{Hz}$ is determined by observation of deliberately excited collective modes of the atom cloud.

\subsection{Control of trap motion}

The transverse movement of the potential is accomplished by applying a time-dependent current to an auxiliary wire running parallel to the main trapping wire.
As shown in the inset of Fig.~\ref{fig:wires}, the additional magnetic field along $z$ causes a slight tilt of the homogeneous bias field, which is exactly aligned along $y$ initially.
The trap minimum position, which is given by the point where the bias field cancels that of the trapping wire, is displaced along $y$.
Additionally, the $y$-component of the modulation field, which changes the magnitude of the bias field, causes a slight proportional movement along $z$.
However, as confirmed by two-dimensional simulations, the anisotropy of the transversal potential suppresses any significant influence on the excitation along $y$.
From numerical simulations of the field geometry, the movement of the trap minimum caused by the current can be calculated as \SI{26}{nm/mA} along $y$ and \SI{9}{nm/mA} along $z$.
The geometry of all chip wires and homogeneous offset fields involved in trapping and modulation is shown in Fig.~\ref{fig:wires}.

\subsection{Effect of finite control bandwidth}
\begin{figure}
\centering
\includegraphics{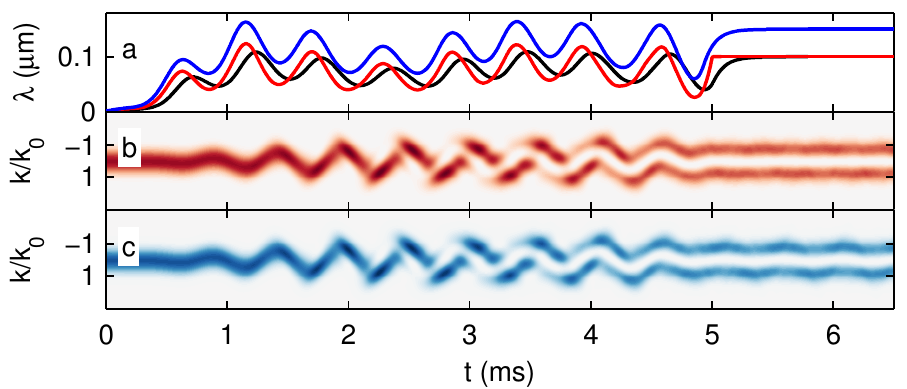}
\caption{
Effect of filtering due to finite electronics bandwidth.
$k$-axes are scaled to $\hbar k_0 = \sqrt{2 m h \nu_1}$
(a) Control ramps $\lambda(t)$. 
Red: original control ramp as derived from OCT. 
Black: control ramp after applying the electronics filtering.
Blue: filtered, rescaled and shifted control ramp.
(b) GPE momentum distribution, simulated without accounting for finite bandwidth.
(c) GPE momentum distribution, simulation including finite bandwidth, rescaling of the control ramp by a factor of 1.6 and a time shift of $\SI{0.08}{ms}$.
}
\label{fig:filter_compare}
\end{figure}

The current in the modulation wire is driven by a custom-design low-noise current source, which is controlled from an arbitrary waveform generator\footnote{Tabor Electronics WW5061}, that outputs the excitation ramp.
A slight smoothing of the control sequence is imposed by finite bandwidth of the electronics, which has to be accounted for when comparing experimental and numerical results (see Sec.~\ref{sec:exp_vs_gpe}).
The measured transfer function modulus $|\mathcal{M}(\nu)|$ at a frequency $\nu$ can be approximated by an exponential $|\mathcal{M}(\nu)| \approx e^{\nu/\nu_\mathrm{co}}$ with cutoff frequency $\nu_\mathrm{co}\approx\SI{4.4}{kHz}$.
Furthermore, a frequency-dependent phase shift is imposed.
Effectively, filtering causes a reduction of the driving amplitude near the resonant frequency~$\nu_\mathrm{1}\approx \SI{1.8}{kHz}$ by a factor $|\mathcal{M}(\nu_\mathrm{y})|^{-1}\sim 1.6$, and a time delay on the order of \SI{0.1}{ms} (Fig.~\ref{fig:filter_compare}d).
In Fig.~\ref{fig:filter_compare} it is shown, that the filtering due to the electronics can be largely canceled by rescaling and shifting the control sequence by these factors.
The difference in the outcome of the simulated momentum distribution is only small and largely given by a slightly enhanced collective oscillation (see Sec.~\ref{sec:shake_two_level}).

\subsection{Measurement of momentum distributions}
\label{sec:detect}

\begin{figure}
\centering
\includegraphics{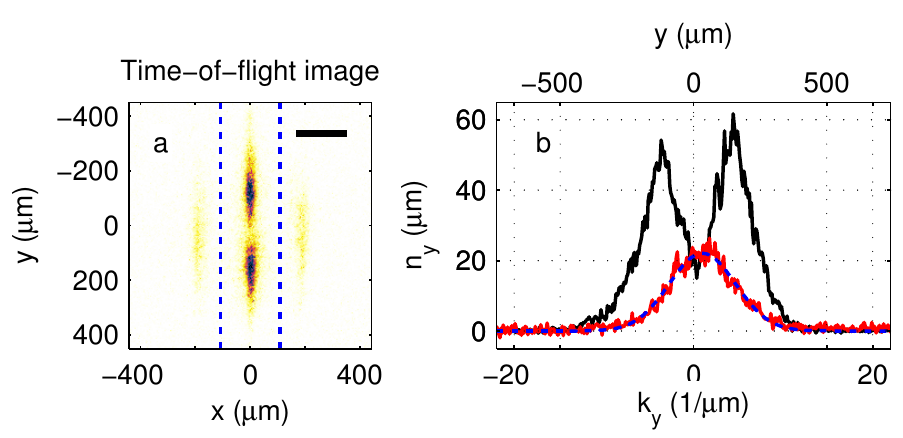}
\caption{
Experimental image analysis. (a) Typical experimental image data for optimal excitation and $t = \SI{5.5}{ms}$, averaged over 12 shots.
As the image is taken after \SI{46}{ms} of expansion time, it predominantly reflects the initial momentum distribution.
The scale bar corresponds to a spatial distance of \SI{187}{\micro m}, equivalent to the typical momentum of $\SI{5.5}{\micro m^{-1}}\approx k_0$.
(b) Transverse momentum distribution inferred from the image in panel (a).
Upper (black) line: central peak, inside dashed lines in (a).
Lower (red) line: emitted atoms, outside dashed lines in (a).
Dashed (blue) line: Gaussian fit to emitted atom momentum.}
\label{fig:image}
\end{figure}

At a time $t$ after starting the excitation process, we suddenly switch off the trapping potential, implying that for $t<\SI{5}{ms}$ the excitation process is still incomplete.
The fast transverse expansion of the cloud due to the tight waveguide confinement causes atom interactions to vanish rapidly, and the ensuing expansion can be considered ballistic.
After $t_\mathrm{exp} = \SI{46}{ms}$ of expansion, we take a fluorescence image~\cite{Buecker2009} similar to that shown in Fig.~\ref{fig:image}(a), fully integrating over the $z$-direction.
In the images, three separate clouds can be observed along the longitudinal $x$-direction. 
The two side peaks emerge due to decay of the excited state which has been populated by our excitation protocol.
Correlation properties of these (twin beams) and the dynamics of the decay process have been analyzed elsewhere~\cite{Buecker2009,Buecker2011}.
We separately integrate along $x$ over the central and side peaks, respectively (blue dashed lines) to analyze the transverse state of each part of the system.
The observed density distributions along $y$ [Fig.~\ref{fig:image}(b)] represent the momentary momentum distributions of the trapped cloud at time $t$, as the initial transverse cloud size (of the order of $l_\mathrm{y}\sim\SI{250}{nm}$) is negligible compared to that after expansion (far field).
If we express momenta as wave numbers $k_\mathrm{y}$, a distance $\delta y$ in the image hence corresponds to $\delta k_y = \alpha \, \delta y$ with $\alpha = m/\hbar t_\textrm{tof} \approx \SI{0.034}{\micro m^{-2}}$.
Taking an experiment series where $t$ is scanned, we can thus fully access the momentum distribution dynamics $\tilde{n}_\mathrm{y}(k,t)$ along the excitation direction, which we will typically depict as false-color plot, see e.g. Fig.~\ref{fig:ioffe_carpet}(a).
Or main interest will be the dynamics of the central (source) cloud which is subject to the excitation; however, in Sec.~\ref{sec:carpet_analysis}, also the transverse dynamics of the twin-beam peaks will be of some importance.
For each of the experimental series shown in this paper, we averaged over typically 12 realizations to suppress noise and allow for robust comparison to theory results.

\section{Results}
\label{sec:results}

We will analyze the excitation dynamics in various complementary ways, motivated by the goal of developing an effective mapping of the many-body dynamics to a driven two-level system.
In Sec.~\ref{sec:exp_vs_gpe} we will start by comparing the results obtained in Sec.~\ref{sec:theory} for the time-dependent momentum distribution of the condensate wave function to experimental observations, varying a range of relevant parameters.
While the excellent agreement ensures that the numerics used to obtain the optimized ramp are accurate, this result only gives limited insight into how the excitation process can be understood qualitatively.
In Sec.~\ref{sec:carpet_analysis} a more phenomenological analysis is performed directly on the experimental data, which will give hints about how to develop a two-level description.
In Sec.~\ref{sec:shake_two_level} the GPE simulations are investigated in more detail, using a description based on Wigner quasi-probability functions, and displaced Fock states.
It will become evident that all approaches lead to conceptually similar and quantitatively compatible interpretations, which can finally be unified to obtain a two-level interpretation as sought after initially.

\subsection{Comparison of experiment and numerics}
\label{sec:exp_vs_gpe}

%

\begin{figure}
\centering
\includegraphics{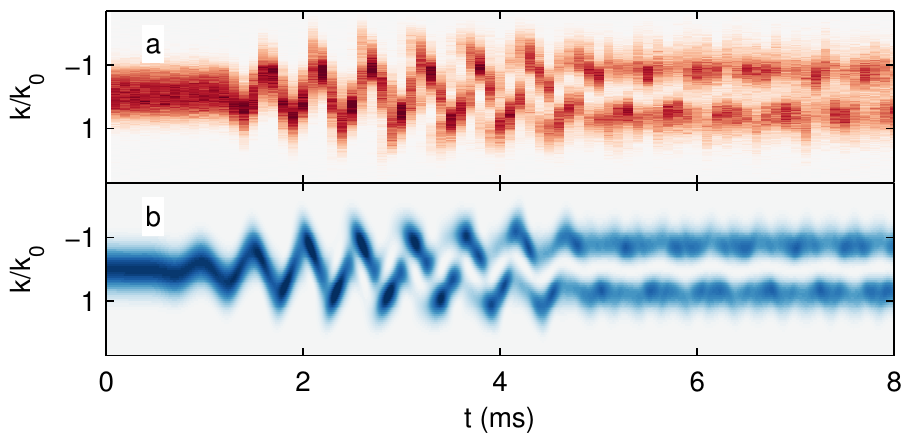}
\caption{
Comparison of momentum distribution dynamics as obtained from experiment and theory for typical parameters.
(a) Experiment.
Each pixel column in the false-color plot corresponds to a distribution as shown in Fig.~\ref{fig:image}(b).
(b) 1d GPE numerics, including finite bandwidth effects (see text).
}
\label{fig:ioffe_carpet}
\end{figure}

Compared to other driven quantum systems, where optimal control techniques may be applicable, a rather unique advantage of cold atoms is the accessibility of the system response, enabled by the relatively large time and length scales and the abundance of powerful imaging techniques.
Probing the performance of a control strategy such as that developed in Sec.~\ref{sec:theory} is not restricted to the final outcome, but the driven system can be monitored even \emph{while} it is being driven, providing direct means to compare experiment and numerical simulations, or apply feedback schemes~\cite{Rohringer2008}.
As explained in Sec.~\ref{sec:detect}, time-of-flight fluorescence images give us direct access to the momentum distribution of the condensate along its transverse axis.
In Fig.~\ref{fig:ioffe_carpet}(a), a typical momentum distribution dynamics plot, as obtained from the experiment, is shown.

\begin{figure}
\centering
\includegraphics{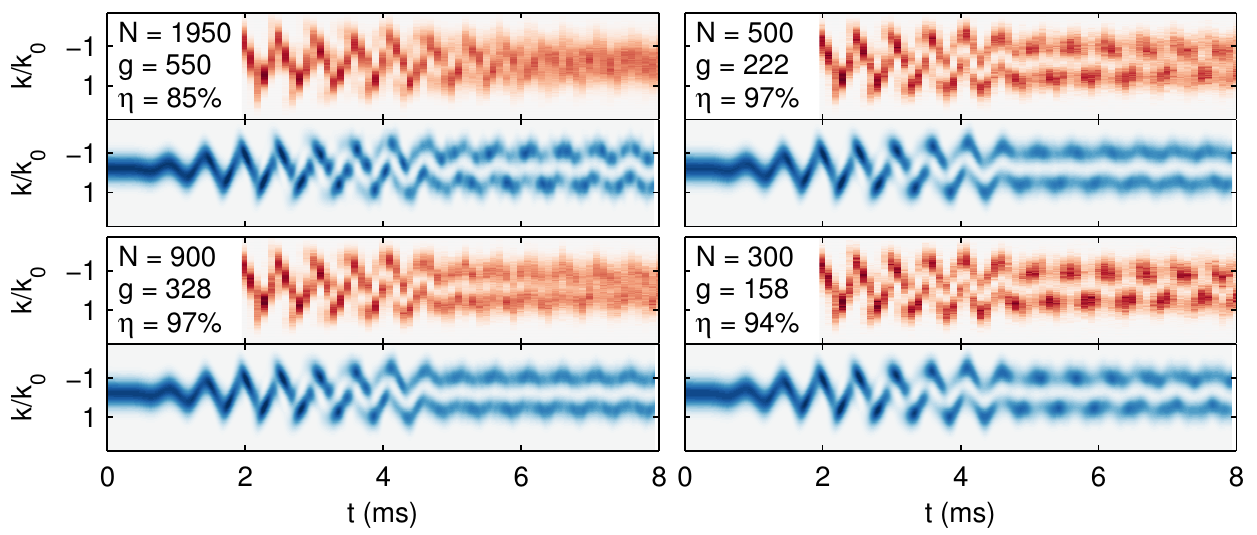}
\caption{Interaction effects on excitation. 
In each pair of plots (top: experiment, bottom: numerics), the typical experimental atom number is shown alongside the GPE nonlinearity term $g$ (in~$\si{Hz \micro m}$, ordinary frequency).
The efficiency $\eta$ is defined as described in the text.}
\label{fig:mean_field_carpet}
\end{figure}

\paragraph{Many-body effects}
Along the transverse directions, confinement is strong enough ($h\nu_\mathrm{1} \gg \mu$) to make interaction-induced effects comparatively small.
Still, to achieve the highest possible fidelity of the excitation, it is crucial to take into account the nonlinear term in Eq.~\eqref{eq:gp} for optimization.
In Fig.~\ref{fig:mean_field_carpet}, the excitation dynamics is shown for a data set, where the atom number has been varied before starting the excitation sequence. 
%
The data is compared to the result of the GPE~\eqref{eq:gp}.

It is observed, that effective excitation is achieved for a nonlinearity corresponding to an atom number $N\sim 900$, which is close to what has been used in the optimization.
For all other atom numbers, stronger residual dynamics after the end of the sequence ($t > \SI{5}{ms}$) is found, indicating decreased fidelity, as the desired state is stationary.
While the GPE simulations reproduce the general tendencies found in the experiment, the agreement is not as good as e.g. for scaled excitations (see below).
For the highest atom number, only rather poor qualitative agreement is reached,  indicating insufficiency of a mean-field model such as GPE (necessitating e.g. a MCTDHB ansatz~\cite{Grond2009,Grond2009b}) and strong effects of the rapid decay of the excited state.

\paragraph{Robustness against experiment inaccuracy}
In OCT, an aspect of high relevance is the sensitivity of the excitation dynamics to deviations of experimental parameters from the ones used for optimization.
In our case, this predominantly applies to parameters affecting the trapping potential.
We consider small changes of the potential parameters $\nu_\mathrm{y}$, $\sigma_\mathrm{y}$. 
In the experiment such deviations arise from variation of the dressing parameters $\Omega_0$, $\Delta_0$, which, in turn, are caused by inaccuracy of the current in the RF antenna wires, and of the external offset field along $x$ (defining the atomic Larmor frequency), respectively.

\begin{figure}
\centering
\includegraphics{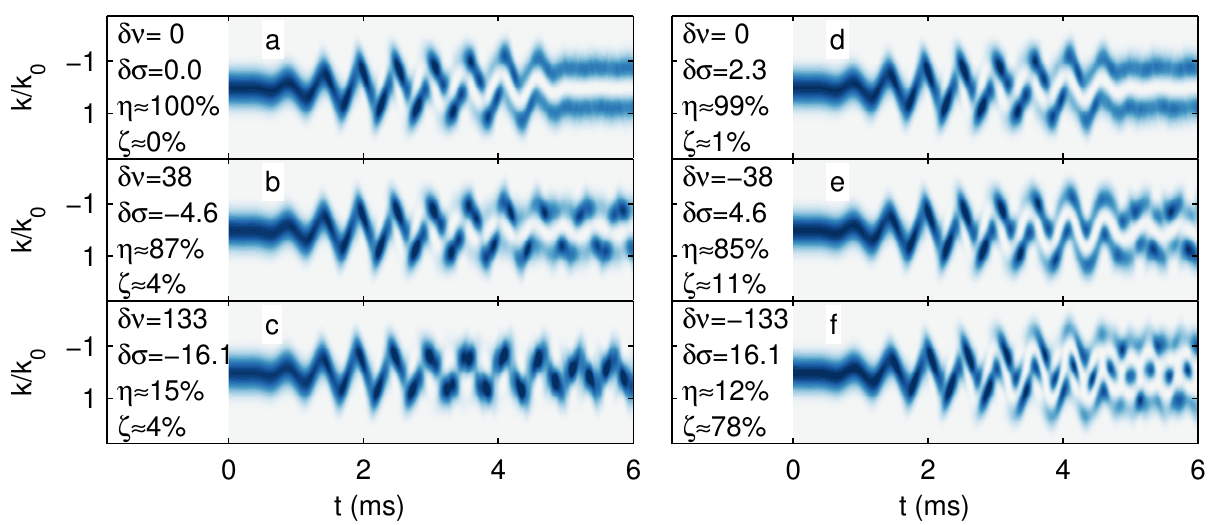}
\caption{
Stability of the excitation sequence against inaccuracy of the trapping potential (numerical result). 
In each plot, the deviation of the potential terms $\delta \nu_\mathrm{y},\delta \sigma_\mathrm{y}$ are given (in units of $\si{Hz}$), as well as the efficiency $\eta$ and the spurious excitation to higher states $\zeta$ as defined in the text.
}
\label{fig:robustness_carpet}
\end{figure}

Numerical results for a range of parameters are shown in Fig.~\ref{fig:robustness_carpet}.
Panels (c-f) correspond to deviations caused by an offset field misalignment of $\pm \SI{2}{mG}$ (b,c) and $\pm \SI{7}{mG}$ (e,f) leading to weaker (positive values) or stronger RF dressing, respectively.
It is observed that any deviation leads to a decrease in excitation efficiency, which is defined here as time-averaged overlap with the desired wave function $\psi_\mathrm{d}$, $\eta = \langle |\braket{\psi_\mathrm{d}| \psi(t)}|^2 \rangle_\mathrm{t > \SI{5}{ms}}$.
The similarly defined population of higher excited states $\zeta$ becomes strong at trap modifications with weaker dressing $\delta \nu_\mathrm{y} > 0$ and $\delta \sigma_\mathrm{y} < 0$.
This effect can be expected, as the protection against excitation to higher states fades with decreasing anharmonicity, while the excursion of the trap relative to the typical length $l_\mathrm{y}$ increases.
In panel (d), on top of an offset field mismatch of $+\SI{3}{mG}$, the current in the RF wire has been adjusted to cancel the effect on $\nu_\mathrm{y}$.
The weak mismatch in $\sigma_\mathrm{y}$ and $\xi_\mathrm{y}$ only leads to a slight reduction of efficiency.
Consequently, optimizing the experimental parameters for a strong excitation (e.g. by minimizing non-stationarity at $t > T=\SI{5}{ms}$) may lead to slightly shifted values, which however compensate.
Using this method, a sensitivity of better than $\SI{1}{mG}$ (or an equivalent mismatch of the dressing current) can be reached, which is beyond what can be achieved by independent characterization of the trapping potential.

\begin{figure}
\centering
\includegraphics[scale=1]{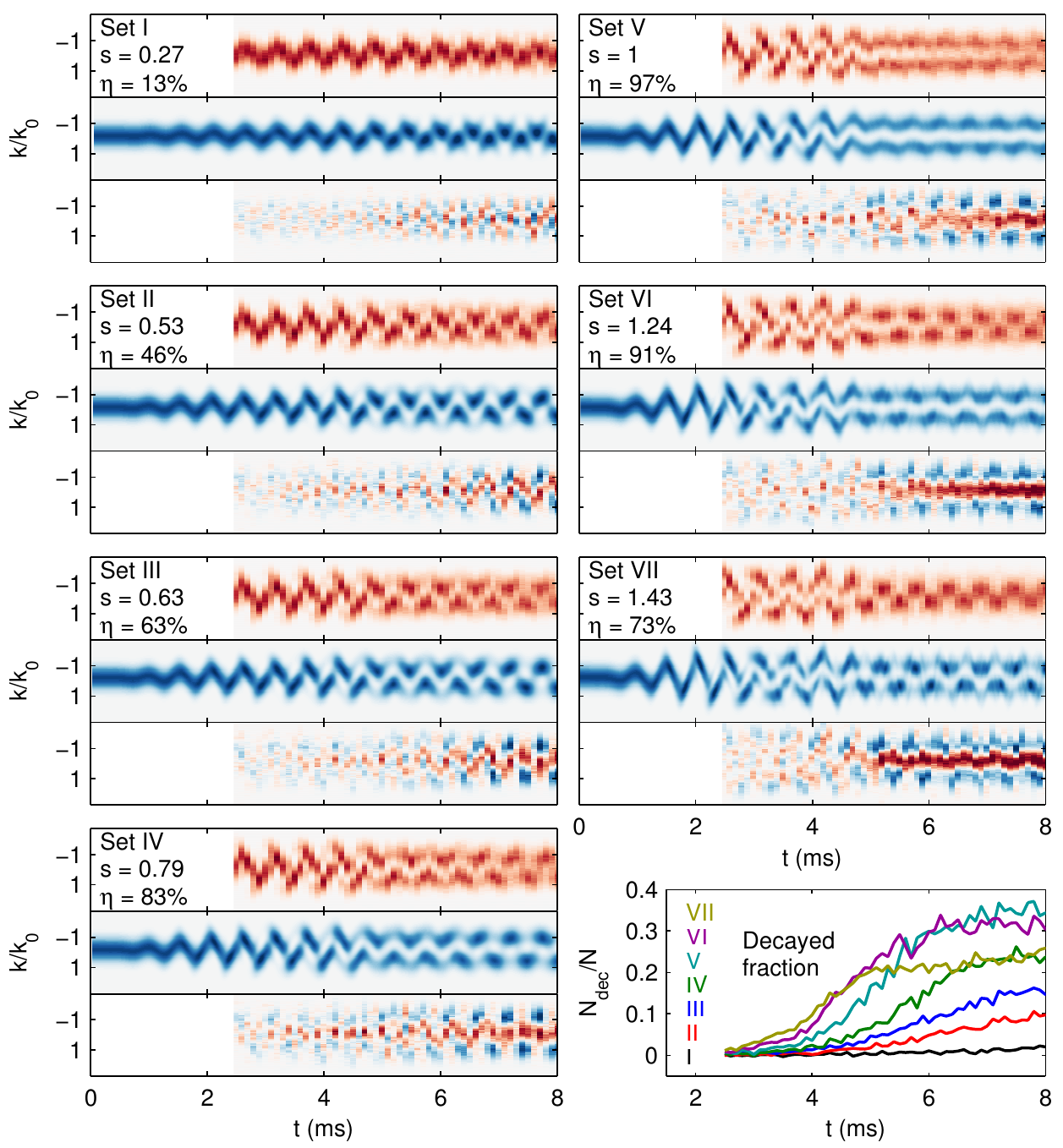}
\caption{Results for scaled excitation ramps.
Mean atom number is $770$ for sets I,II,IV-VI, and $856$ for sets III and VII.
For each of the seven sub-sets, the upper image (red false-color)  is the experimental result, normalized separately for each time step.
The middle image (blue false-color) shows the numerical GPE result, including low-pass filtering and scaling by the factor $s$ as given.
The bottom image shows the deviation between experiment and theory, expressed as imbalance $\tilde{n}_\mathrm{ex} - \tilde{n}_\mathrm{th}$; the color scale for the imbalance is enhanced by a factor 3.
The bottom right inset shows the relative amount $N_\mathrm{dec}/N$ of atoms that have decayed from the excited state into twin atom pairs.
}
\label{fig:gpe_compare_big}
\end{figure}

\paragraph{Scaled excitations}
In Fig.~\ref{fig:gpe_compare_big} the momentum distribution dynamics is shown for a data set with varying excitation efficiency, which will be the main subject of analysis in the remainder of this and the following section, as it covers a very broad range of control sequences.
To achieve different efficiencies, the excitation ramp has been scaled in amplitude by factors $s$ with respect to the optimal control result, resulting in strongly varying wave function dynamics.
The approach of simple amplitude scaling has been chosen over using separately optimized ramps for different efficiencies, to allow for easier comparison due to the well-defined relation between the used control sequences. 
Furthermore, our analysis will show that the main spurious effect of this strategy are collective oscillations at reduced scalings. 
Comparison between GPE and experimental result (average over $\sim 12$ realizations) shows excellent agreement at early times.
\footnote{Note that $s$ has been defined including the necessary re-scaling due to finite electronics bandwidth (see Fig.~\ref{fig:filter_compare}).
}
At later times, decay of the excited state into twin beams, which is not accounted for in theory, becomes significant (see bottom right panel) and for high values of $s$, agreement is reduced due to inelastic collisions with the twin beams which reside in a different transverse state.
However, for weak excitation, even the shape of single ``beating peaks'' after the end of the excitation pulse is precisely captured by numerics.
Along the $k$-axis, the GPE result has been convolved with a Gaussian of $m/(\hbar t_\mathrm{tof}) \cdot \SI{40}{\micro m} \approx \SI{1.20}{\micro m^{-1}}$ rms width to account for finite imaging resolution and bulk position fluctuations. 
Apart from a small shift of the $t$-axis and a slight re-scaling of the $k$-axis,\footnote{The shift in $t$ is well below the experimental time resolution, and is very probably due to the inaccuracy of the filtering circuit characterization. 
The necessity for the re-scaling of $k$ (of the order of 10\%) might arise from interaction effects causing weak hydrodynamic effects in expansion~\cite{Kruger2010}.
The values of both adjustment are consistent among all sets shown.} the scaling factor $s$ is the only free input parameter of the simulation.

Having established the detection method, and verified that the outcome is consistent with the numerics on which the control optimization has been founded, we now proceed to a more qualitative analysis of the experimental result.

\subsection{Analysis of experimental momentum dynamics}
\label{sec:carpet_analysis}

In this and the following section we will analyze the momentum distribution dynamics beyond a simple comparison to numerical results.
The notion underlying the discussion will be that of a few-level system, comprised by the ground, first and occasionally second excited state of the confinement potential along the excitation direction, with the final goal to reduce the anharmonic oscillator to a closed two-level system.\footnote{In the literature on quantum control strategies, this problem is occasionally discussed as that of leakage-suppression of a two-level system~\cite{Motzoi2009,Rebentrost2009,Khani2009,Jirari2009}.}
This approach may seem inappropriate, as it relies on the superposition principle, which requires a linear equation of motion and is hence not applicable to a mean-field wave-function as described by the GPE.
However, in our case the nonlinearity is weak compared to the oscillator energy, and so is the modification of the dynamics due to many-body effects (see Fig.~\ref{fig:mean_field_carpet}), suggesting that a description in terms of single-particle states may still provide significant insight.

\begin{figure}
\centering
\includegraphics[scale=1]{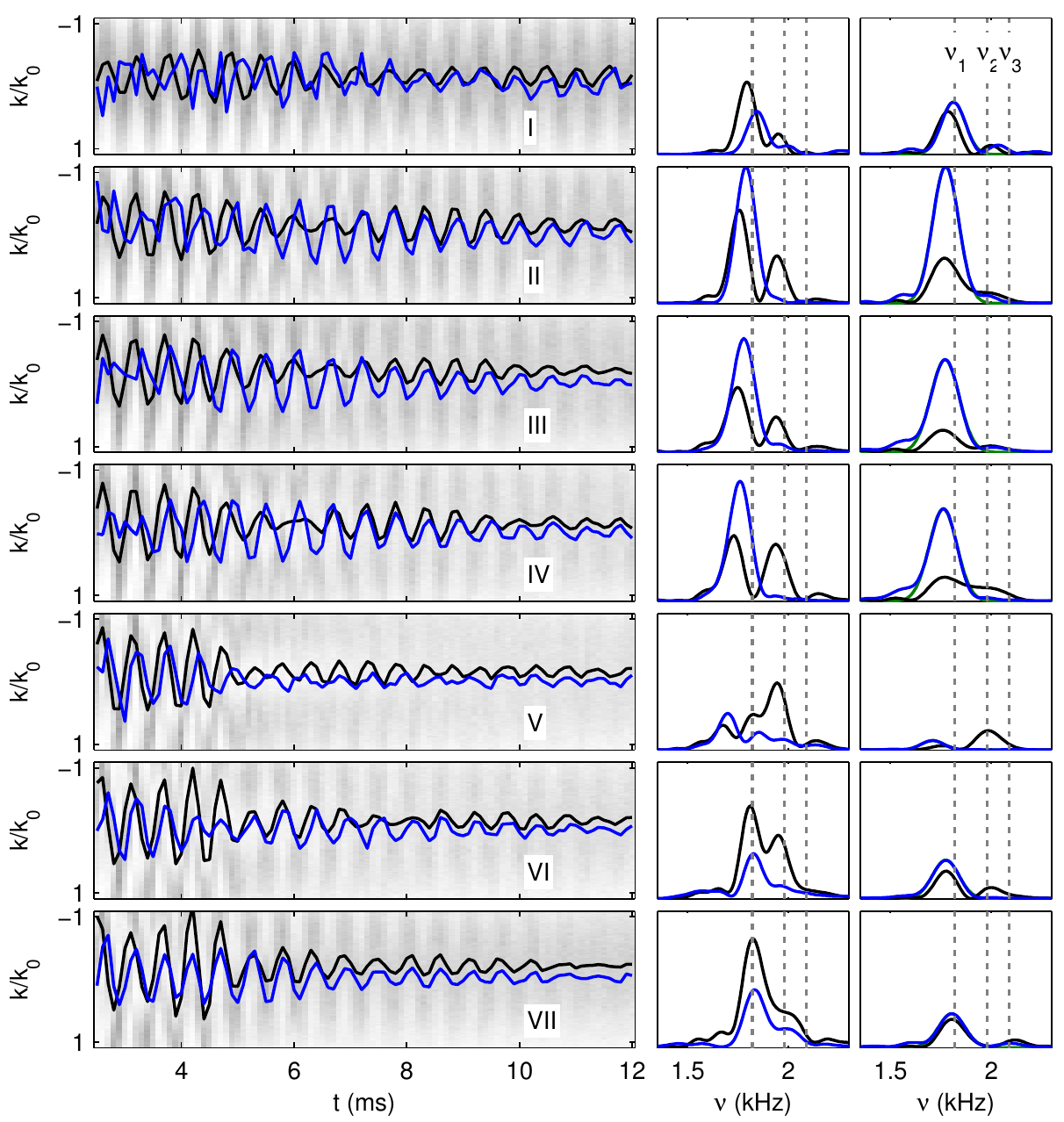}
\caption{
Momentum space center-of-mass dynamics for data set as shown in Fig.~\ref{fig:gpe_compare_big}.
Left column: center-of-mass momentum of the source cloud with respect to a fixed frame ($K(t)$, black) and relative to the twin-beam center-of-mass ($K_\mathrm{r}(t)$, blue).
(See Fig.~\ref{fig:frame_compare} for the twin-beam center-of-mass.)
In the background, the full dynamics is shown (see Fig.~\ref{fig:gpe_compare_big}).
Middle column: corresponding power spectra $f(\nu)$, taken over the entire time span shown.
Right column: spectra, taken over a time span starting from $t > T=\SI{5}{ms}$, i.e., after the end of the excitation.
Grey dashed lines in the background indicate the harmonic frequency $\nu_\mathrm{h}$, and the first three level spacings, as defined in the previous section.
All spectra are in arbitrary units, but normalized identically for each of the columns.
}
\label{fig:spectra}
\end{figure}

\paragraph{Center-of-mass dynamics}
As the simplest possible observable derivable from the momentum dynamics, we start by analyzing the transverse center-of-mass of the experimental images, corresponding to the momentum expectation value $K(t) \equiv \braket{k_\mathrm{y}(t)}$, see black lines in Fig.~\ref{fig:spectra} (left panels).
In the power spectra of $K(t)$ (center panels), two strong peaks are observable near the first two transverse level spacings at frequencies $\nu_1 = \SI{1.84}{kHz}$ and $\nu_2 = \SI{1.99}{kHz}$, and a weak third at $\nu_3\approx\SI{2.10}{kHz}$, defined analogously.
Assuming a single-particle level picture, these peaks can be interpreted as beating frequencies between populations of the first three levels of the oscillator, where mean-field effects are causing frequency shifts, as described below.
Consequently, the magnitude of oscillations is the strongest for intermediate excitation efficiencies (sets II-IV), where the levels are populated most evenly, maximizing the beating contrast (see below).

A crucial observation is, that also the transverse profiles of the twin-beam peaks, which are separated in the images longitudinally (see Fig.~\ref{fig:image}), exhibit strong oscillations of $K_\mathrm{t}(t) \equiv \braket{k^\mathrm{(t)}_\mathrm{y}(t)}$. 
Meanwhile, they fully maintain their Gaussian shape [Fig.~\ref{fig:image}(b)].
In Fig.~\ref{fig:spectra}, oscillations of the \emph{relative} center-of-mass $K_\mathrm{r}(t) = K(t) - K_\mathrm{t}(t)$  (left), and their power spectrum (center) $f(\nu) = |\mathcal{F} [K_\mathrm{r}(t)](\nu)|^2$, are shown as blue lines.
It is observed that, while the oscillations are similarly strong as in a fixed frame, all peaks in the power spectrum, except that near $\nu_1$ are suppressed.
This suggests, that in a reference frame co-oscillating with $K_\mathrm{t}(t)$, the dynamics can be understood in terms of two transverse levels, motivating an approach of decomposition into a quasi-classical oscillation, and ``internal'' dynamics, which remain unaffected by the bulk oscillation.\footnote{This decomposition is exactly valid for harmonically confined many-body systems~\cite{Garcia-Ripoll2001,Bialynicki-Birula2002}.
Obviously, this does not hold for an anharmonic oscillator, which is exactly why our excitation to a non-classical state by displacement can work at all (see above). 
Being aware of the inconsistency, we still apply the decomposition approach to qualitatively understand the dynamics.}
This interpretation is consistent with our understanding of the decay process~\cite{Buecker2012}, where the transverse state of the twin beams, a ground state displaced by $K_\mathrm{t}(t)$, defines the appropriate ground state for the internal dynamics.
In Sec.~\ref{sec:shake_two_level}, a more rigorous formalism for the co-oscillating frame will be given, and its position will be independently derived from numerical results.

In the right column of Fig.~\ref{fig:spectra}, spectra are shown which are derived from the oscillations at times $t > \SI{5}{ms}$ only, i.e., where no driving occurs anymore.
Hence, they provide a characterization of the final state that is reached after the excitation.
Qualitatively, the same features are observed as in the full time spectra, however, peaks at $\nu_2$ are smaller, which is consistent with theory, as will be shown below.
Also, in the relative center-of-mass spectrum, the observation of a single-peak structure, with a minimal amplitude for the most efficient excitation is even more evident.

\begin{figure}
\centering
\includegraphics[scale=1]{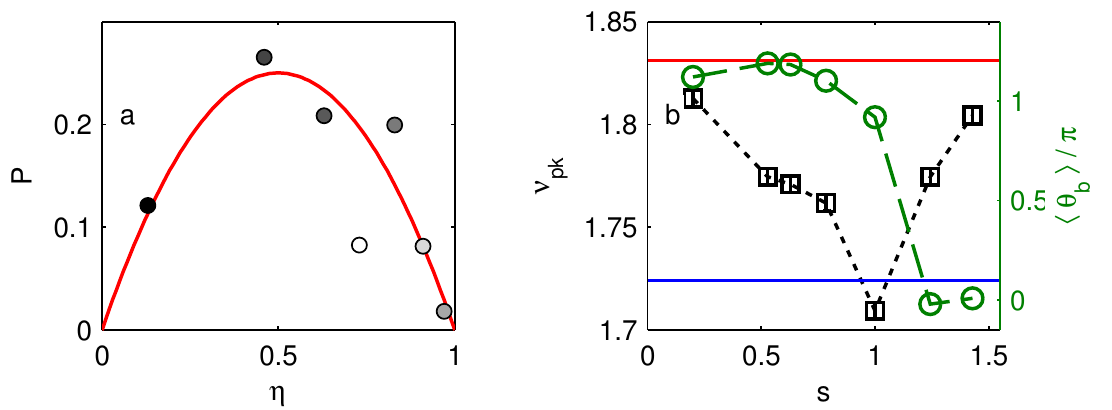}
\caption{Analysis of post-excitation beating spectra shown in the right column of Fig.~\ref{fig:spectra}. 
(a) Integrated power of oscillations $P$. 
The experimental points have been scaled along the $y$-axis for best fit to $\eta (1-\eta)$ (red line).
$\eta$ has been derived as described in the previous section.
The shading of each point indicates the corresponding scaling $s$ (white is highest).
(b) Peak position (black, left axes) and cosine of averaged phase (green, right axes).
Red and blue lines correspond to the single-particle level spacing $\nu_1$, and the mean-field-shifted level spacing $\nu_1'$, respectively.}
\label{fig:spectra_peak}
\end{figure}

In Fig.~\ref{fig:spectra_peak}(a), the integrated power of the oscillations $P \propto \int f(\nu) \mathrm{d}\nu$, measuring the stationarity of the final state, is shown as a function of the numerically obtained excitation efficiency $\eta$ (see previous section).
Apart from the strongest driving, where higher states may become excited more easily, $P$ shows fair agreement with a curve given by $\eta (1-\eta)$, which is the squared amplitude of the interference term in the momentum-space density of a two-level system with momentum-space wave functions $\tilde{\psi}_0,\tilde{\psi}_\mathrm{d}$:
\begin{align}
\label{eq:beating_fun}
\tilde{n}(k_\mathrm{y},t;\eta) &= \left| \sqrt{1-\eta} \tilde{\psi}_0(k_\mathrm{y}) + \sqrt{\eta} \tilde{\psi}_\mathrm{d}(k_\mathrm{y}) \right|^2 \\
&= (1-\eta) |\tilde{\psi}_0(k_\mathrm{y})|^2 + \eta |\tilde{\psi}_\mathrm{d}(k_\mathrm{y})|^2\nonumber  \\
&\qquad + 2\sqrt{\eta(1-\eta)}\Re[\tilde{\psi}_0^*(k_\mathrm{y})\tilde{\psi}_\mathrm{d}(k_\mathrm{y})]\cos(2\pi\nu'_1 t). \nonumber
\end{align}

The positions of the beating peak (obtained from a Gaussian fit) are shown in Fig.~\ref{fig:spectra_peak}(b).
For high efficiency, the frequency is shifted downwards from the oscillator level spacing $\nu_1$ (red line).
This is explained by the mean field term in the GPE~\eqref{eq:gp}.
For the boundary case of near-unity efficiency, the shift can be calculated rather easily. 
As the ground state population is negligible, it does not contribute to the interaction energy, and the chemical potential $\mu_\mathrm{e}$ for an atom in the excited state is given by the second eigenvalue of the time-independent GPE.
The according wave function $\psi_\mathrm{d}$ (i.e., the desired state in the optimization process) can now be used to calculate the chemical potential of a single atom in the ground state $\psi'_0$, using a Schrödinger equation with effective potential arising from the mean field of the excited state:
\begin{align}
\mu_\mathrm{e} \psi'_0(y) = \left[ -\frac{\hbar}{2m}\frac{\partial^2}{\partial y^2} + V_\mathrm{ext}(y) + 2 g |\psi_\mathrm{d}(y)|^2 \right] \psi'_0(y).
\end{align}
The beating frequency is now given by the difference in chemical potential. 
Instead of the oscillator level spacing $\nu_1 \approx \SI{1.831}{kHz}$, we obtain $\nu_1' = (\mu_e - \mu_g)/(2\pi\hbar) \approx \SI{1.724}{kHz}$ (blue line).
Given the uncertainty in the input parameters of the calculation (such as the assumption of an equilibrium Thomas-Fermi shape longitudinally), this value agrees well with the experimentally obtained one for maximum efficiency (set IV), $\nu_\mathrm{V} = \SI{1.709(4)}{kHz}$.

Finally, we can have a look at the phase of the (relative) center-of-mass oscillation.
When comparing the value of $K^\mathrm{(r)}_\mathrm{y}(t)$ for different scalings at a fixed time in Fig.~\ref{fig:spectra}, it is apparent, that the phase inverts at the point of maximum efficiency.
We take the averaged phase from the Fourier transform result, weighted by the Lorentzian fit of the peak, and obtain the curve shown in Fig.~\ref{fig:spectra_peak}(b, right axes).
The inversion is reminiscent of a two-level system subject to a Rabi driving, where the phase of precession inverts after passing the pole of the Bloch sphere at a pulse area larger than $\pi$.
As will be shown in Sec.~\ref{sec:shake_two_level}, the excitation process can be understood analogously.

\subsection{Interpretation of numerical result: two-level driving model}
\label{sec:shake_two_level}

\begin{figure}
\centering
\includegraphics[width=\textwidth]{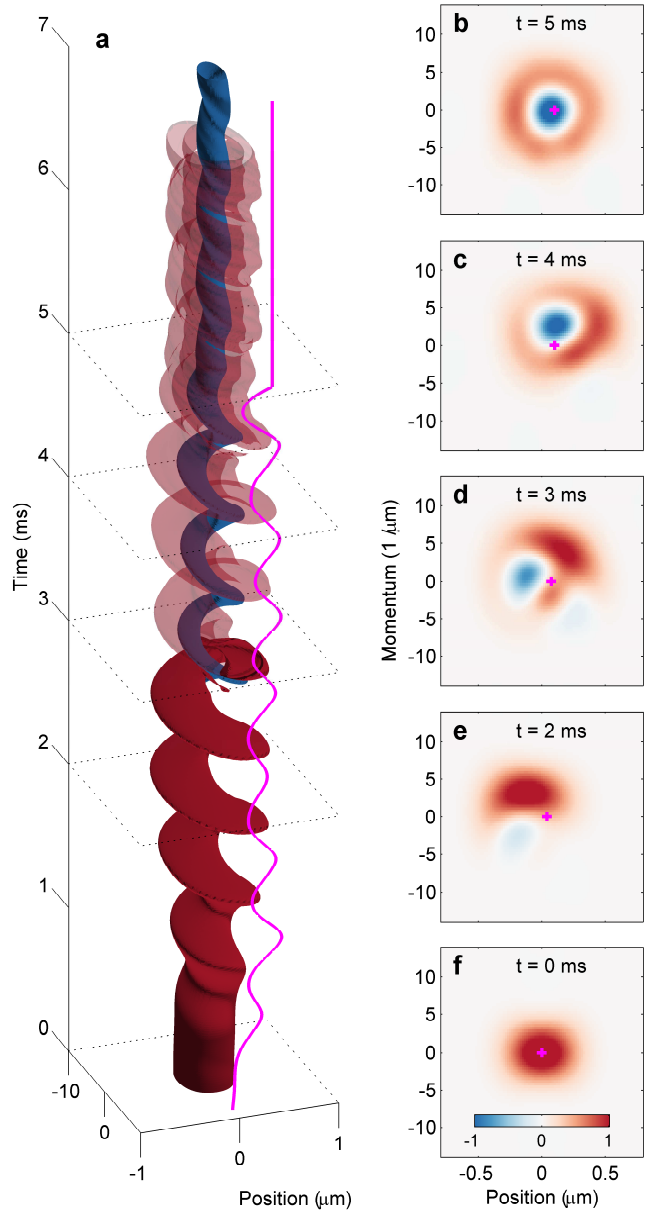}
\caption{Time evolution of Wigner function.  Panel (a) reports the time evolution of the Wigner function, and panels (b--f) show snapshots at selected times.  In (a) we show the iso-surfaces at $\pm 0.35$ times the maximum value of the Wigner function, with transparency added to the iso-surface at time above 3~ms to highlight the appearance of the negative Wigner function part, associated with the first excited state.  For discussion see text.}
\label{fig:wigner}
\end{figure}

To understand the physical mechanism governing the optimal excitation protocol, in the following we analyze the time evolution of the condensate wave function in the Wigner representation~\cite{Walls2007}:
\begin{equation}\label{eq:wigner}
  W(y,k,t)=\int e^{-iks}\psi(y+\frac s 2,t)\psi^*(y-\frac s 2,t)\,\mathrm{d}s\,,
\end{equation}
which provides a mixed position-momentum distribution.  The Wigner function has many appealing features reminiscent of a classical distribution function.  Integration over all momenta $k$ gives the spatial probability distribution $|\psi(y,t)|^2$.  Likewise, integration over $y$ gives the momentum probability distribution.  The Wigner function of the condensate ground state, Fig.~\ref{fig:wigner}(f), approximately corresponds to the ground state of the harmonic oscillator, with equal uncertainty in position and momentum.  In the figure the distribution is slightly elongated along $y$ due to the nonlinear atom-atom interactions.  The desired state of the control, Fig.~\ref{fig:wigner}(b), corresponds to the first excited state of the GPE in the anharmonic trap.  It has positive and negative values (giving a node at $y=0$ upon integration over all momenta), and thus differs from a genuine classical distribution function.

Panel (a) of the figure reports the time evolution of the Wigner function.  We plot the iso-surfaces at $\pm 0.35$ times the maximum value of the Wigner function.  At times later than 3 ms we have added transparency to the iso-surface for positive values to show the appearance of the negative part of the Wigner function, associated with a non-classically excited state.  The solid line shows the time variation of the spatial minimum of the confinement potential. Initially, this displacement brings the condensate into collective oscillations, whose frequency is determined by the harmonic part of the confinement potential. For a large enough displacement, the condensate wave function is brought into the region where the anharmonicity of the confinement is sufficiently large to modify the internal structure of the wave function (and not just its displacement).  One observes that in addition to the center-of-mass oscillations in this regime the transfer from the ground to the first excited state occurs.  Finally, at the terminal time $T=5$ ms of the control process the minimum of the confinement potential is shifted to bring the condensate to a complete halt.

We next suggest a procedure to approximately map the excitation dynamics onto a genuine two-level description of ground and excited condensate states.  
As in Sec.~\ref{sec:carpet_analysis}, the main idea is to separate the wave function dynamics into (i) a collective, quasi-classical oscillation, which is needed to bring the condensate into the anharmonic part of the trap, and (ii) an internal conversion between the ground and first excited state, defined in a co-moving frame.
The latter conversion is governed by the anharmonic part of the trap, as explained above. 

\begin{figure}
\centering
\includegraphics[scale=1]{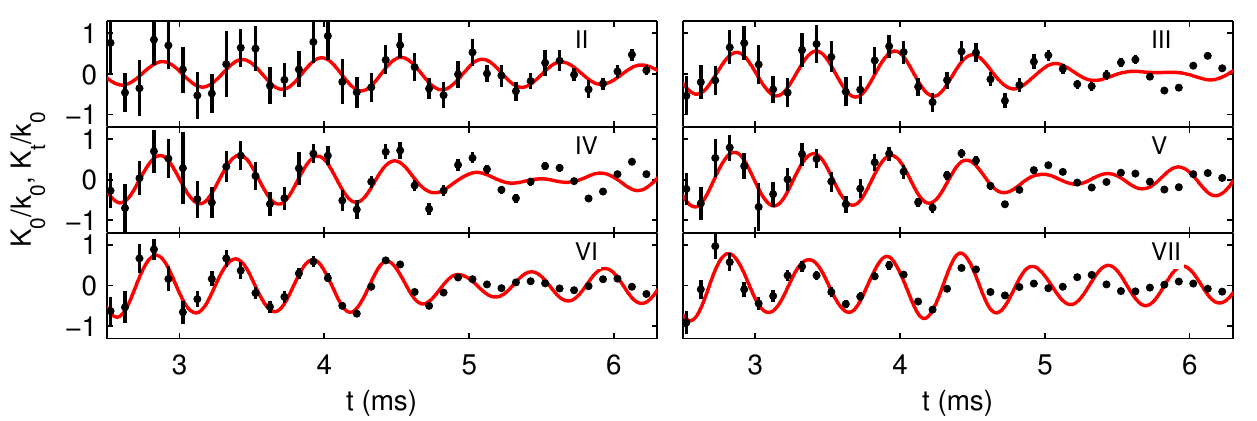}
\caption{Reference frame for two-level model.
Underlying data are the same as shown in Figs.~\ref{fig:gpe_compare_big} and~\ref{fig:spectra}.
Red lines are the momentum-space displacement $K_0(t)$ of the two-mode basis states, as obtained from applying Eq.~\eqref{eq:two_mode_overlap} to the GPE result.
Black points indicate the experimentally found center-of-mass position of twin beams that have decayed from the excited state $K_\mathrm{t}$, defining the reference frame for the emission process (see Sec.~\ref{sec:carpet_analysis}).
Similar to the momentum space dynamics as shown in Fig.~\ref{fig:gpe_compare_big}, agreement reduces at later times, where decay into twin beams becomes strong.
Data set I has been omitted due to the emission of twin beams being insufficient to determine $K_\mathrm{t}$.
}
\label{fig:frame_compare}
\end{figure}

\begin{figure}
	\includegraphics{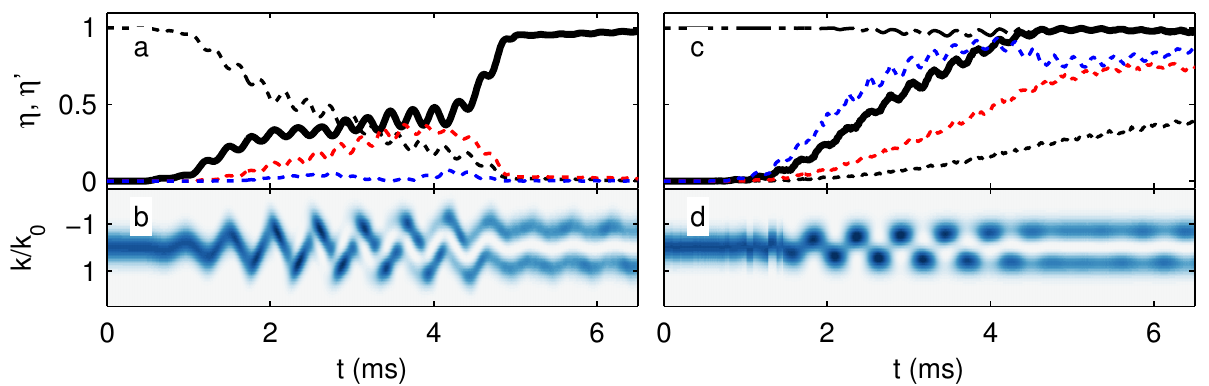}
	\caption{State populations during the excitation process. (a) Populations $\eta(t)$ of excited states of the full anharmonic potential [Eq.~\eqref{eq:potential6_initial}] as arising from direct projection of the GPE result for data set V ($s=1$).
	The solid line indicates the population of the first excited state, dotted lines represent the ground (black) and first and second excited (red, blue) states.
	(b) Corresponding momentum dynamics (identical to Fig.~\ref{fig:gpe_compare_big}-V).
	(c) Population of first excited state in co-oscillating frame within the two-mode model $\eta'(t)$.
	Solid line: data set V, corresponding to solid line in (a).
	Dashed lines: sets I (black, $s=0.27$), III (red, $s=0.63$), and VII (blue, $s=1.43$).
	The dash-dotted line indicates the total overlap of the two-level model with the GPE result $\chi(t)$ [see Eq.~\eqref{eq:two_mode_overlap}], which exceeds a value of 0.95 at all times $t$.
	(d) Momentum dynamics arising from time-dependent superposition of $\phi_0, \phi_1$ in the co-oscillating frame, data set V.
	Note the strong beating at intermediate times/excited fractions.
	}
	\label{fig:populations}
\end{figure}

We define wave functions $\phi_\mathrm{g}(y)$ and $\phi_\mathrm{e}(y)$ as single-particle eigenfunctions of the \emph{harmonic} part of the trap potential only, i.e. Eq.~\eqref{eq:potential6_initial} with $\sigma_\mathrm{y}$ and $\xi_\mathrm{y}$ set to zero.
Also, any modifications due to the nonlinear atom-atom interactions are neglected. 
This simplification allows us to analyze the dynamics in terms of displaced Fock states~\cite{DeOliveira1990}, that capture well the notion of the separation approach.
Let $\hat{D}[\alpha(t)] = \exp [\alpha(t)\hat{a}^\dagger-\alpha(t)^*\hat{a}]$ denote the displacement operator of the harmonic oscillator~\cite{Walls2007}, where $\alpha(t)= [ l_\mathrm{y}^{-1} Y_0(t)+ i l_\mathrm{y} K_0(t)]/\sqrt{2}$ determines the position and momentum of the displacement at time $t$, and $\hat{a}$ denotes the annihilation operator.
For a given displacement $\alpha(t)$, we can compute the overlap between the displaced ground and excited states with the condensate wave function according to
\begin{equation}
\label{eq:two_mode_overlap}
\chi(t) = \left|\int\left[\hat{D}[\alpha(t)] \phi_g(y)\right]^*\psi(y,t)\,\mathrm{d}y\right|^2+
  \left|\int\left[\hat{D}[\alpha(t)] \phi_e(y)\right]^*\psi(y,t)\,\mathrm{d}y\right|^2\,.
\end{equation}
Determining the value $\alpha(t)$ which gives the largest overlap at time $t$ allows us the aforementioned decompositions into (i) center-of-mass coordinates $Y_0(t)$ and $K_0(t)$, and (ii) probability amplitudes $\langle\hat{D}[\alpha(t)] \phi_g|\psi(t) \rangle$ and $\langle\hat{D}[\alpha(t)] \phi_e|\psi(t) \rangle$ for the ground and excited state within the displaced frame.  
In all cases we find an overlap $\chi(t)$ well above 90\%, which thus justifies the wave function decomposition.
In Fig.~\ref{fig:frame_compare}, the obtained values for $K_0(t)$ are shown as red lines, and compared to experimentally obtained values, as described below.
In Fig.~\ref{fig:populations} we compare $\chi(t)$ and the obtained excited population $\eta'(t) = [\langle\hat{D}[\alpha(t)] \phi_g|\psi(t) \rangle|^2$ to results from direct projection of $\psi(y,t)$ on the oscillator states $\phi(y)$ (defined in the co-moving frame of the excitation motion). 
The direct projection leads to strong transient population of higher excited states, and a sudden jump near the end of the excitation [Fig.~\ref{fig:populations}(a)], where they are depopulated again.
This is reminiscent of the fixed-frame center-of-mass spectra (black lines in Fig.~\ref{fig:spectra}), where a peak near $\nu_2$ is present when regarding the entire sequence (center column), but mostly vanishes after $t=T$ (right column).
In contrast, the two-level approximation in the system displaced by $\alpha(t)$ yields a smooth transition [Fig.~\ref{fig:populations}(b)], consistent with the continuous appearance of negative values of the Wigner function (Fig.~\ref{fig:wigner}).
In the momentum dynamics derived from the two-level model in a similar manner to Eq.~\eqref{eq:beating_fun} as shown in Fig.~\ref{fig:populations}(d), a continuous transfer to the excited state is observed, with strong beating at intermediate excited population.
Again, this is consistent with the experimental \emph{relative} center-of-mass spectra (blue lines in Fig.~\ref{fig:spectra}), where only a single peak near $\nu_1$ persists, even during the excitation.
Similar to a Rabi pulse with area larger than $\pi$, the excited population $\eta(t)$ is decreasing towards $t=T$ for scaling parameters $s > 1$.

As laid out in Sec.~\ref{sec:carpet_analysis}, the appropriate ground state for the internal conversion dynamics can also be determined in the experiment from the center-of-mass position $K_\mathrm{t}(t)$ of the twin-atom beams which the excited state is decaying into continuously.
For times $t$, where the decayed fraction becomes perceivable, we can compare the experimentally found $K_\mathrm{t}(y)$ to $K_0(y)$ as in Fig.~\ref{fig:frame_compare}, and find good agreement without any free parameter over a large range of settings.
Together with the absence of decay products from higher excited states in the experiment, this result confirms the validity of the decomposition approach.
In Ref.~\cite{Buecker2012} it has been shown, that the obtained populations of the excited state lead to an accurate quantitative description of the ensuing decay process.

\section{Conclusion}
In conclusion, we have presented successful application of optimal control theory to the problem of preparing a non-classical, strongly out-of-equilibrium motional state of a Bose-Einstein condensate, realizing population inversion with near-unity fidelity.
The obtained condensate wave function corresponds to the first excited eigenstate of the Gross-Pitaevskii equation, closely resembling the first odd Fock state of a harmonic oscillator.
To manipulate the external state of the Bose-Einstein condensate, we used precisely controlled motion of an anharmonic trap potential along the optimized trajectory.
Experimental and numerical results on the momentum distribution dynamics during and after the excitation sequence show excellent agreement over a large range of parameters, including tuning of many-body effects.
Moreover, a model of the excitation dynamics based on decomposition into a quasi-classical oscillation and the actual state transfer has been developed, and shown to be consistent with various observations made in both experiment and theory.
Using this approach, we were able to deduce an approximate two-level description of the excitation process.

A first application of the vibrational state inversion, using the condensate as a gain medium for matter wave amplification, has been demonstrated in Refs.~\cite{Buecker2011,Buecker2012}.
However, optimal control in condensates is not restricted to high-fidelity preparation of a desired wave function, and more general pulses that e.g. acting on non-stationary initial states in a phase-sensitive manner can be implemented~\cite{VanFrank2013}.
State preparation beyond a mean-field description has been proposed, including entanglement generation~\cite{Platzer2010,Caneva2012}, number-squeezed states~\cite{Grond2009}, or cooling~\cite{Rahmani2012}, which should be realizable in a similar fashion.
More generally, our results highlight the potential of experiments with Bose-Einstein condensates as a test-bed for a large range of quantum control problems, as known from NMR spectroscopy~\cite{Skinner2003, Skinner2004, Khaneja2005} solid-state,~\cite{Hofheinz2008,Hohenester2006,Safaei2009,Jirari2009, Rebentrost2009,Khani2009}, atomic,~\cite{Vasilev2009}, or molecular physics~\cite{Assion1998, Schirmer2001,Tesch2002}.

\section*{Acknowledgments}
We acknowledge financial support from the Austrian Science Fund projects P24248, CAP, SFB FoQuS, the FWF docotoral program CoQuS (W 1210), and the EU project AQUTE.

\section*{References}


\end{document}